\documentclass{article}

\usepackage[utf8]{inputenc}  
\usepackage[T1]{fontenc}     
\usepackage{hyperref}        
\usepackage{graphicx}        
\usepackage{amsmath, amssymb} 
\usepackage{xcolor}          
\usepackage{pgfplots}        
\pgfplotsset{compat=1.16}
\usepackage{subcaption}      
\usepackage{natbib}          
\usepackage{caption}         
\usepackage{booktabs}        
\usepackage[utf8]{inputenc}

\title{Why Does the Cortex Have Such a Vast Storage Capacity?}

\author{
    Hui Wei\thanks{Laboratory of Algorithms for Cognitive Models, Fudan University, Shanghai 200433, China. Corresponding author: weihui@fudan.edu.cn} \and
    Surun Yang \and
    Yangwang Li
}
\date{}

\begin{document}

\maketitle

\begin{abstract}
The capacity of long-term memory seems to be extremely large, capable of storing information spanning almost a lifetime. Why does it have such a vast capacity? Why are some memories so enduring? What is the actual physical form of long-term memory? In the movie Inside Out, it is depicted as individual orbs containing information. Is that really the case? Simply explaining this by saying that the cortex has many neurons, numerous neural connections, and complex electrochemical activity between them is not sufficient to answer these fundamental questions. We need to uncover the theory hidden behind these phenomena.In essence, a neural network is equivalent to a very large directed graph, with a massive number of nodes and directed connections. This paper posits that the physical form of long-term memory is a connected subgraph within this complex directed graph. This subgraph is capable of linking together the disparate fragments of the same event, spread across different sensory cortices, to form associations. This provides a physical realization of the engram theory. The robustness of the connected subgraph and the resources it consumes can explain various memory behaviors.Based on anatomical, brain imaging and electrophysiological evidence, this paper constructs a probabilistic connectivity model and uses theorems from graph theory to prove the ease of constructing connected subgraphs. Finally, it explains why the potential capacity for memory is immense.
\end{abstract}


\section{Introduction}
Although the amount of information stored in a 1TB hard drive is completely clear, the amount of information stored in a piece of cortex is quite obscure. Physical representations of digital formats as photos or text prove to be hard drive or the floating gate of a solid-state drive which are, by so far, understood and delicately designed for information carriage. Quantifying storage for a brain remains unclear if we dive deep into questions like "What is the storage format of a piece of information in the brain?", "How much resource does storing a single piece of information consume?", "How much total resource does the brain have?", "How are these resources allocated?" and the alike. When referring to mechanical formats, every single process how they supplement information transformation is comprehended, key detail for the process of the physical implementation form of semantic memory or episodic memory remains misty.

Storage, consolidation and retrieval being its core functions, memory system prove an information processing system which retains information hence to alter future action modes. Storage refers to the formation of memories, ensuring future effective retrieval through encoding. Consolidation strengthens the durability of memories, assuring persistence for delayed retrieval. Retrieval involves reactivating stored information through clues, thus enabling recall or recognition. Speculations about the framework of the memory system have been made based on behavioral tests, to establish the multi-store memory model proposed by Atkinson and Shiffrin\cite{atkinson1968human}.

Episodic memory being a representative long-term memory, various sensory channels act as its components like visual, auditory, tactile, olfactory, gustatory, and spatial senses to store event-related information. When specific cue is activated, the entire context regarding the event is recalled. In research on context-dependent memory, Baddeley and Godden differentiated the contexts between the learning and recall phases in their tests. Experiment results indicated that participants performed better during the recall phase when they were positioned in a context similar to that of the learning phase\cite{godden1975context}. This suggested that a context similar to the learning phase promotes information retrieval and that the multiple aspects of an event are conjunctly encoded. How is this mechanism implemented? The Contextual Binding Model accounts for an explanation, which hypothesizes that when a memory is first processed by the hippocampus, the areas responsible for recognizing "what" the object is and "where" it is are connected to the hippocampus\cite{yonelinas2019contextual}. However, this explanation is still too vague when to consider the details how memory is physically represented.

In addition to behavioral science, neuroscience has also conducted meticulous researches on memory, the engram theory being one of the current hotspots. The term "engram" was first proposed by Richard Semon in 1911, originally referring to changes in the brain induced by external stimuli, which lead long-lasting effects \cite{semon1925mneme}. In modern neuroscience experiments, engram transformation process is considerably represented at somatic level. Classic studies relating engram cells suggest that initial engram forms in hippocampus and gradually transport to the mPFC (medial Prefrontal Cortex) during consolidation. When long-term memory engrams are concreted, during the reactivation phase, neuronal signals travel through the mPFC-BLA (basolateral amygdala) pathway to elicit corresponding somatic receptor. For instance, in contextual fear conditioning, the shock response of mice to a conditioned stimulus proves identical with BLA engram activity \cite{josselyn2020memory, kitamura2017engrams,tonegawa2018role}. Experiments analyzing the concept of engram as entire ensembles have been studied, discovering that when a specific engram pathway is activated, various subregional ensembles are activated, and similar activations occur during slow wave sleep \cite{roy2022brain,wang2023functional}. It is suggested that the presence of a specific engram is not only physically represented by particular ensemble in the hippocampus or BLA, but rather spreads to different parts of the entire brain.
The hippocampus, BLA or mPFC ensembles may serve as hubs that trigger the activation of the whole engram map, allowing one to recall stored details \cite{roy2022brain,wang2023functional}. When compared to the connections between non-engram cells, the synapses between engram cells exhibit morphological and functional specificity, which may prove morphological representation of memory retrievability\cite{ChoiDongIl2021Scoa,Hayashi-TakagiAkiko2015Laoe,lee2023hippocampal,lee2023neocortical}. These collective biological evidence urges a plausible indicative theoretical framework. 

A powerful analytical toolchain provided by Marr’s theory suits to view memory as such information processing system \cite{MarrD.1982Vaci}. An information system is suggested to be clarified with three levels of illustration, computation, algorithm and implementation. The conclusions of behavioral experiments align with the computational level, demonstrating the functional features of memory. Neurobiological methodologies like MRI, EEG, somatic or molecular analysis may serve as physical implementation. The algorithm remains unclear, serving as the missing link between the macro and micro levels.

Directed graphs in graph theory may serve as an intermediate level apparatus, to conjunct neuronal details and psychological outcomes. This parallel distributed network may be presumed as the algorithmic implementation, thus complex electrochemical reactions between neurons as interactive activities between the nodes of directed graphs. Engrams hence might be interpreted as a connected subgraph within a directed graph, which representatively specifies an unique event. 
Memory encoding appears the formation of such a connected subgraph, strengthening of synapses during consolidation in accordance with strengthening of directed edges, to lift subgraphical stability, retrieval relating to reactivation of subgraphs stored previously. This approach fundamentally differentiate from using graphs to study functional connectivities between cortical regions, in which nodes are regarded cortical or subcortical regions, edges as event-related connectivities, graph algorithm as relevance revealed.

In graph analysis of brain networks, five generative undirected graph models are frequently utilized: ER random graphs, WS small-world networks, scale-free networks, spatial models, and stochastic block models\cite{lynn2019physics}.

Spatial models site the actual spatial location of neurons and fail to reflect the dynamic interactions and information transference between different brain regions. In contrast, small-world networks, ER random graphs, scale-free networks, and stochastic block models exhibit significant structural differences compared to data gathered from clinical experiments. In reference to in-degree out-degree issues speaking of undirected graph models, it is hypothesized that the connection probability between neurons decreases exponentially as distance increases.\cite{horvat2016spatial,perin2011synaptic,pillai2012dendritic}. Meanwhile, Kramer compared undirected graph generation models with clinical experimental data on amygdala neurons, and found significant discrepancies between the two \cite{kramer2023spatial}.

To conclude, to develop a directed graph network generation method that aligns with neurobiological details, with directed subgraph physically representing engrams, and analysis of reachability, discrimination, and coexistence of these connected subgraphs supplies plausible estimate of memory capacity.

\section{Results}
\subsection{ Memory engrams are essentially connected subgraphs}
 In neuroscience, neurons that are activated and involved in the storage of a specific memory during the process of memory formation are referred to as Engram Cells. Representing an event often requires a large number of cells, such as those widely distributed across different sensory channels, and these dispersed cells are connected through certain connecting units. This is analogous to a connected subgraph in graph theory. If we abstract the neural network into a highly connected, decentralized, and active directed graph, different Engram Cells would represent the nodes in the graph. When these nodes are connected, they form a connected subgraph within the vast directed graph of the brain. This connected subgraph could be the physical realization of a memory engram. Multiple connected subgraphs can coexist within the network, share some of the same nodes, and may lose parts of their branches due to resource competition. This physical realization can comprehensively explain the neurobiological implementation of the encoding, consolidation, and retrieval processes. It can also account for the conclusions drawn from psychological memory experiments.
\subsection{ Probabilistic model of cortical neuron connections}
\begin{figure*}[thbp]
\centering
    \begin{subfigure}[b]{0.4\textwidth}
        \centering
        \begin{tikzpicture}
            \begin{axis}[
                xlabel={Distance Between Two Neurons},   
                ylabel={Probability of Connections},   
                enlargelimits=false, 
                axis on top,         
                legend pos=north east, 
                width = 5.5cm,
                height = 5.5cm,
            ]
            \addplot[thick,color={rgb:red,0.17;green,0.627;blue,0.172}] graphics[xmin=0, xmax=315, ymin=0, ymax=0.55] 
            {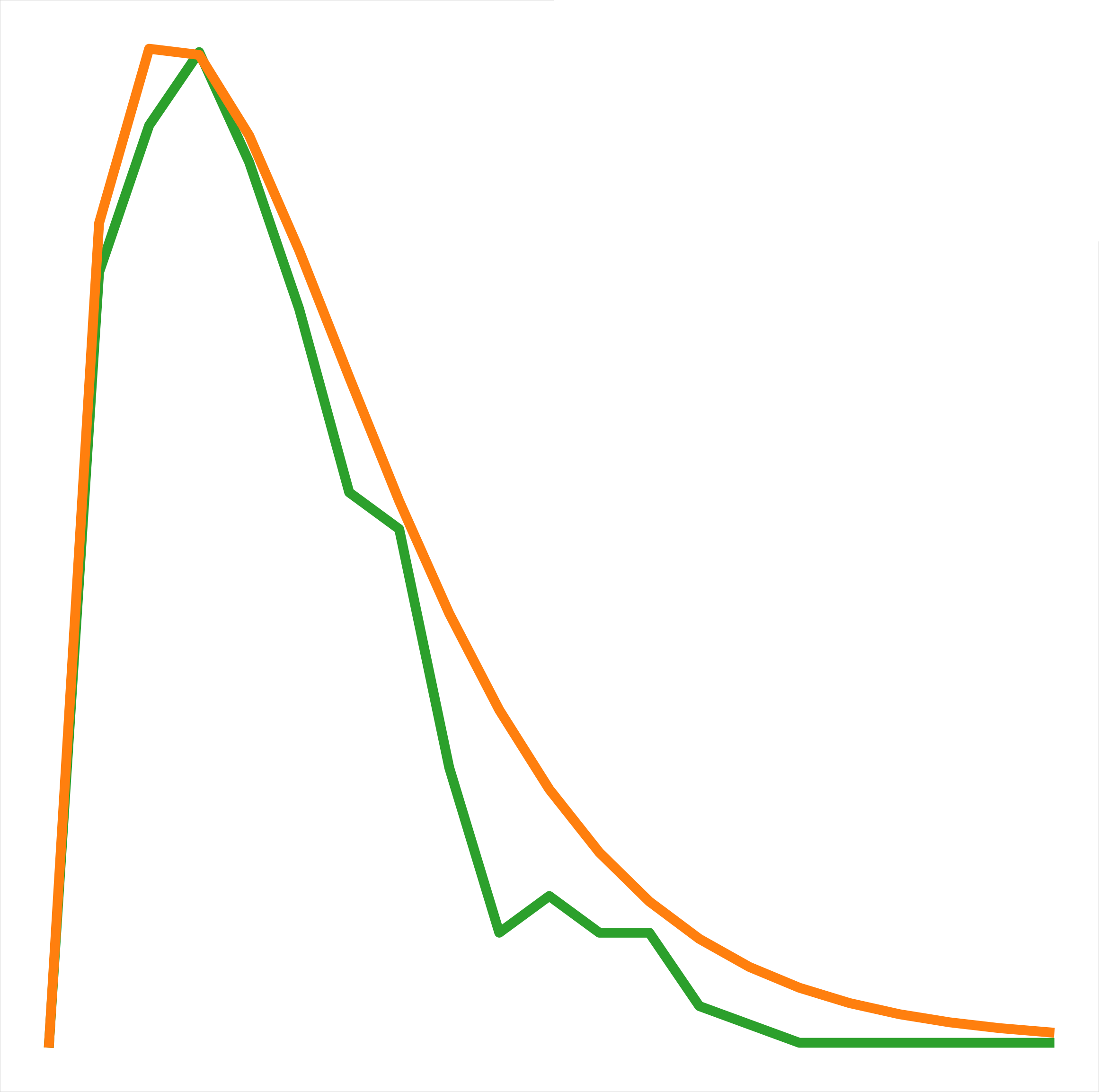};  
            \addplot[thick,color={rgb:red,0.99;green,0.49;blue,0.054}] graphics[xmin=0, xmax=315, ymin=0, ymax=0.55] 
            {pics/figure1-1};  
            \addlegendentry{EXP\cite{budd2001local}} 
            \addlegendentry{our model} 
            \end{axis}
        \end{tikzpicture}
        \caption{$k$=0.06,$\lambda$=41, compares the results from Budd}
        \label{fig:Budd}
    \end{subfigure}
    \hspace{0.03\textwidth}
    \begin{subfigure}[b]{0.4\textwidth}
        \centering
        \begin{tikzpicture}
            \begin{axis}[
                xlabel={Distance Between Two Neurons},   
                ylabel={Probability of Connections},   
                enlargelimits=false, 
                axis on top,         
                legend pos=north east, 
                width = 5.5cm,
                height = 5.5cm,
            ]
            \addplot[thick,color={rgb:red,0.17;green,0.627;blue,0.172}] graphics[xmin=0, xmax=950, ymin=0, ymax=0.9] 
            {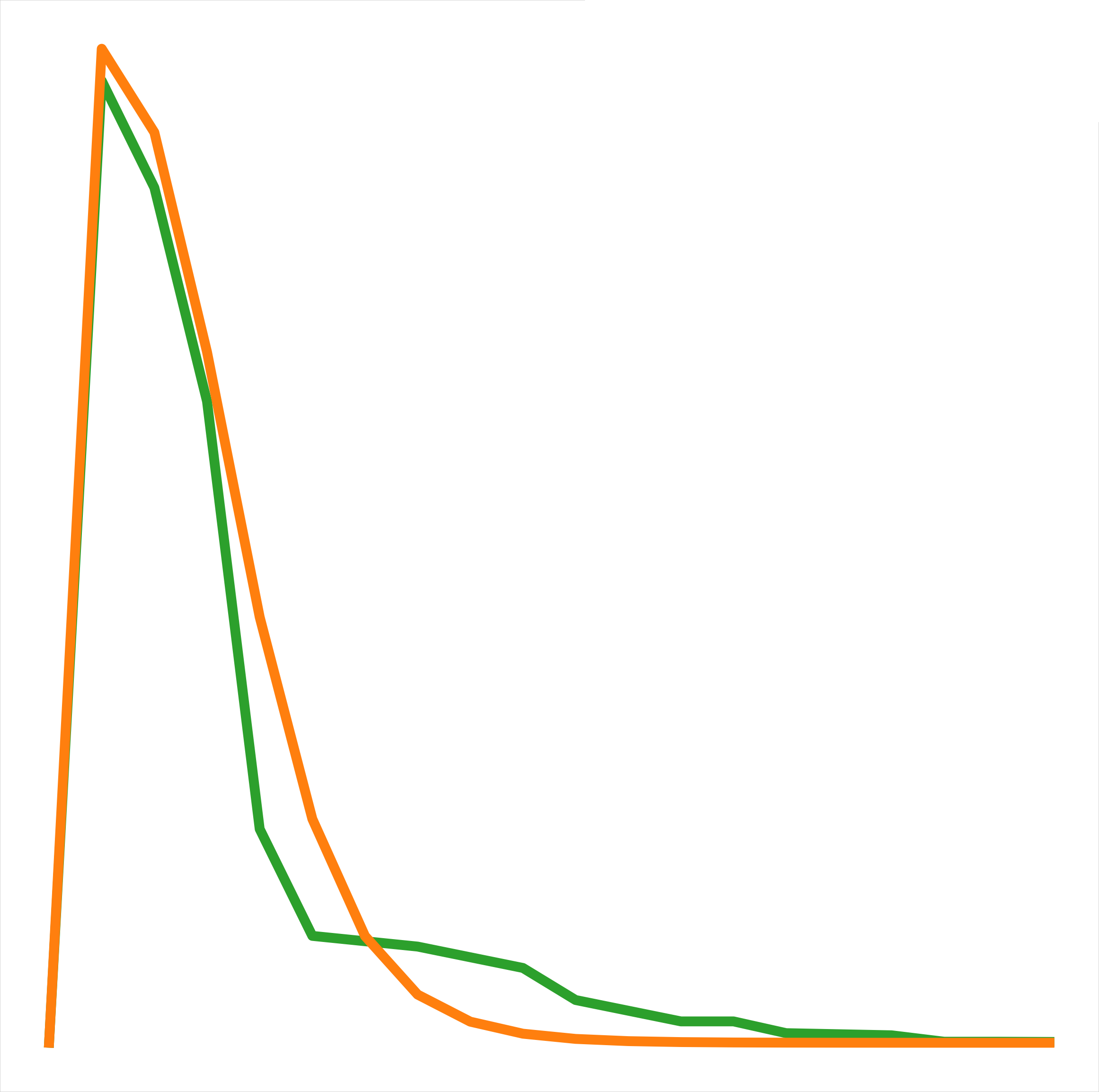};  
            \addplot[thick,color={rgb:red,0.99;green,0.49;blue,0.054}] graphics[xmin=0, xmax=950, ymin=0, ymax=0.9] 
            {pics/newfigure1-2};  
            \addlegendentry{EXP\cite{horvat2016spatial}} 
            \addlegendentry{our model} 
             \end{axis}
        \end{tikzpicture}
        \caption{$k$=0.15,$\lambda$=55, compares the results from Horvát}
        \label{fig:Horvat}
    \end{subfigure}
    \caption{The function describing the connection probability between neurons, with the yellow curve representing our model and the green curve representing the experimentally measured data.}
\end{figure*}

 The connectivity properties of directed graphs must be supported by neurobiological experimental evidence. Anatomically, there is an observation that short-range connections between neurons are more common than long-range connections. However, this is not precise enough to construct large-scale directed graphs that approximate real neural networks. Biological neural networks are not trivial, unconstrained directed graphs, and the connectivity properties of such graphs must accurately reflect the objective reality of neural networks, at least statistically. To equate a neural network to a very large directed graph, we first need to establish the connection rules between each node in the graph. If the formation of synaptic connections by neurons at a specific location in space is considered an independent event, and if two neurons have enough synapses within a sufficiently small neighborhood, then it is possible for these two neurons to form synaptic connections within that neighborhood\cite{tecuatl2021comprehensive}. Based on these anatomical and electrophysiological findings, we derived a probabilistic model of neuron connections that reflects biological reality and verified that the model parameters are consistent with anatomical evidence. The connection probability between two neurons is a probabilistic function related to the distance between them. Through derivation (for details, see the SI Material and Methods: Neuron Connection Probability Modeling), the number of synapses $N(d)$ formed between two neurons at a distance $d$ can be expressed as a function of several parameters: the distance d between the two neurons, the distance $r$ from the synapse formation site to the soma, a constant $ A $ (representing the proportion of upstream and downstream synapses that are likely to connect within a very small neighborhood), the synaptic density $\rho_0$ at the center of the ellipsoid, and the synaptic density decay rate $\lambda$. This yields the following equation \ref{eq:Nd}:
 \begin{equation}
    N \left(d\right) = \mathsf{A} \rho_{0}^{2} \sum e^{- \frac{\sqrt{2 \left(\right. r^{2} + \left(d - r\right)^{2} \left.\right)}}{\lambda}} \Delta r \textrm{ }
    \label{eq:Nd}
\end{equation}
The connection probability between two neurons is determined by the number of synapses between them. Research has shown that the Poisson distribution can be used to fit the relationship between the potential number of synapses and the connection probability between neurons\cite{kersen2022connectivity} . The probability that a pair of neurons forms x synaptic connections is given by:
\begin{equation}
    P \left(x = n\right) = \frac{\left(N \left(d\right)\right)^{n}}{n !} e^{- N \left(d\right)}
\end{equation}
In reality, multiple synaptic connections may exist between two neurons. For example, the upstream axon terminal can form connections with various parts of the downstream neuron, such as the soma, the distal or proximal dendrites, or the axon hillock. When constructing a directed graph model in this paper, these multiple synaptic connections are aggregated into a single edge. Therefore, the connection probability between two neurons is the probability that there is at least one synaptic connection between them, which is the probability of being non-zero:
\begin{equation}
    \begin{split}
    P \left(be\_connected\right) &= P \left(x \neq 0\right) = 1 - P \left(x = 0\right)\\
    &= 1 - \frac{\left(N \left(d\right)\right)^{0}}{0 !} e^{- N \left(d\right)}\\
    &= 1 - e^{- N \left(d\right)}\\
    &= 1 - e^{- \mathsf{A} \rho_{0}^{2} \sum e^{- \frac{\sqrt{2 \left(r^{2} + \left(\left(d - r\right)\right)^{2}\right)}}{\lambda}} \Delta r \textrm{ } \textrm{ }}\\
    \end{split}
\end{equation}
Let $k = \mathsf{A} \rho_{0}^{2}$,
\begin{equation}
P \left(be\_connected\right) = 1 - e^{- k \sum e^{- \frac{\sqrt{2 \left(r^{2} + \left(d - r\right)^{2}\right)}}{\lambda}} \Delta r \textrm{ }}
\label{eq:beconnected}
\end{equation}
Next, need to determine which set of parameters can generate a directed graph that aligns with biological reality. This paper select a directed graph with 500 nodes for subsequent experiments (the reason for choosing 500 nodes can be found in the SI Material and Methods: Modeling Directed Graphs that Meet Biological Constraints). By adjusting different values of $k$ and $\lambda$,  various connection probabilities between cells observed in different biological experiments can be simulated. As shown in \ref{fig:Budd} compares the results with the data from Budd et al.\cite{budd2001local}, \ref{fig:Horvat} compares the results with the data from Horvát et al.\cite{horvat2016spatial}. These comparisons demonstrate the consistency between the theoretical parameters and actual experimental data. This consistency validates the parameters, which can then be used for further simulations of directed graph modeling.

\begin{table}
\centering
\caption{Model Parameter}
\begin{tabular}{l p{5cm}}
Parameter & Meaning \\
\midrule
$d$ & the distance between the two neurons \\
$r$ & the distance from the synapse formation site to the soma \\
$\rho_0$ & the synaptic density at the center of the ellipsoid\\
$A$ & the proportion of upstream and downstream synapses that are likely to connect within a very small neighborhood\\
$\lambda$ & the synaptic density decay rate\\
$N \left(d\right)$ & the number of synapses formed between two neurons at a distance $d$\\
$P \left(X = n\right)$ & The probability that a pair of neurons forms $X$ synaptic connections\\
\bottomrule
\end{tabular}
\end{table}

\subsection{ The mathematical guarantee of engrams is assured}
A connected subgraph can serve as the physical implementation of an engram, provided that a directed graph has sufficient capacity, meaning it can accommodate a sufficient number of coexisting compatible connected subgraphs. Additionally, neurons activated by a particular event are dispersed throughout the cortex, and they  should be connected with a relatively high success rate. For a set of nodes, if multiple connectivity schemes exist, then they can be easily connected. In graph theory, a Hamiltonian cycle refers to traversing all nodes in a graph and returning to the starting point. This theory can also be applied to a subset of nodes in a directed graph.\cite{hefetz2016random, ferber2018counting} proved that for a random directed graph $D(N,p)$, where $N$ is the number of nodes and $p$ is the independent probability of an edge existing (with at most $N\left(N-1\right)$ edges), as long as $D(N,p)$ satisfies the Asaf-Ferber condition:
The in-degree and out-degree of any node are both non-zero, and $p>\log{N}/{N}$, then:

Hamiltonian cycles must exist, and the number is typically $N!\left(p\left(1+o\left(1\right)\right)\right)^N$.

\begin{figure*}[t]
    \centering
    \begin{minipage}{0.7\textwidth}  
        \centering
        \includegraphics[width=\linewidth]{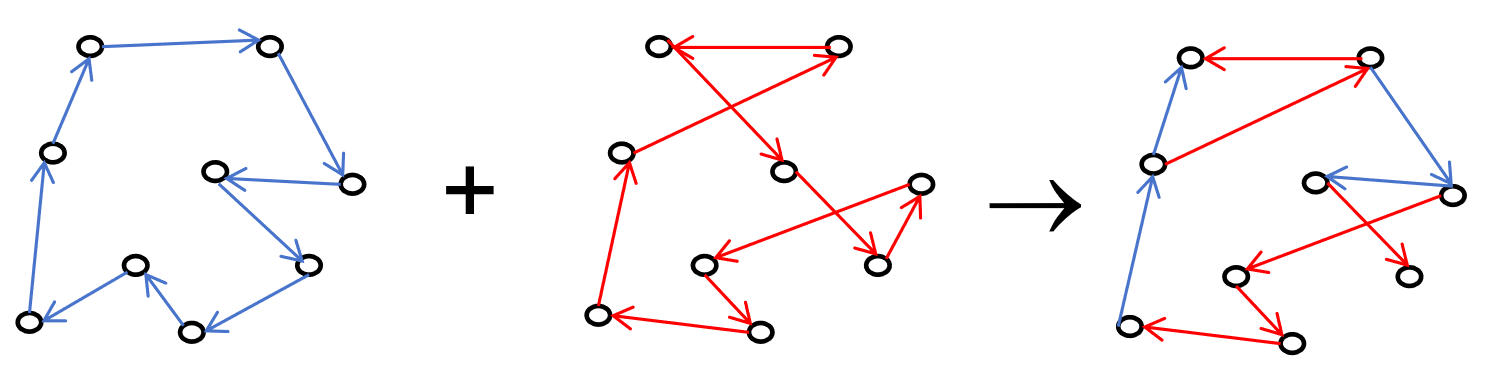}  
        \caption{If a subset of nodes contains multiple Hamiltonian cycles, then connected subgraphs can be easily obtained.}
        \label{fig:multipleHamiltonian}
    \end{minipage}
\end{figure*}

A Hamiltonian cycle has stricter constraints than trivial connectivity, and this theorem characterizes the theoretical lower bound of graph connectivity, indicating the minimum that can be achieved. If a subset of nodes has more than one Hamiltonian cycle, it is very easy to ensure they are simply connected, as shown in \ref{fig:multipleHamiltonian}. If a directed graph's subgraph contains a sufficient number of Hamiltonian cycles, it indicates that activating a portion of the nodes in this subgraph can provide multiple stable pathways for the awakening and recall of the whole subgraph, which aligns very well with the initial motivation of engram theory.

\begin{figure*}[thbp]
    \centering
    \begin{subfigure}[b]{0.4\textwidth}
        \centering
        \begin{tikzpicture}
            \begin{axis}[
                xlabel={Scale of Number of Subgraph Nodes},   
                ylabel={Average Connection Saturation},   
                ylabel style={yshift=-1em},
                enlargelimits=false, 
                axis on top,         
                legend pos=south east, 
                width = 6cm,
                height = 6cm,
            ]
            \addplot[red, dashed, thick] graphics[xmin=0, xmax=420, ymin=0, ymax=0.7] 
            {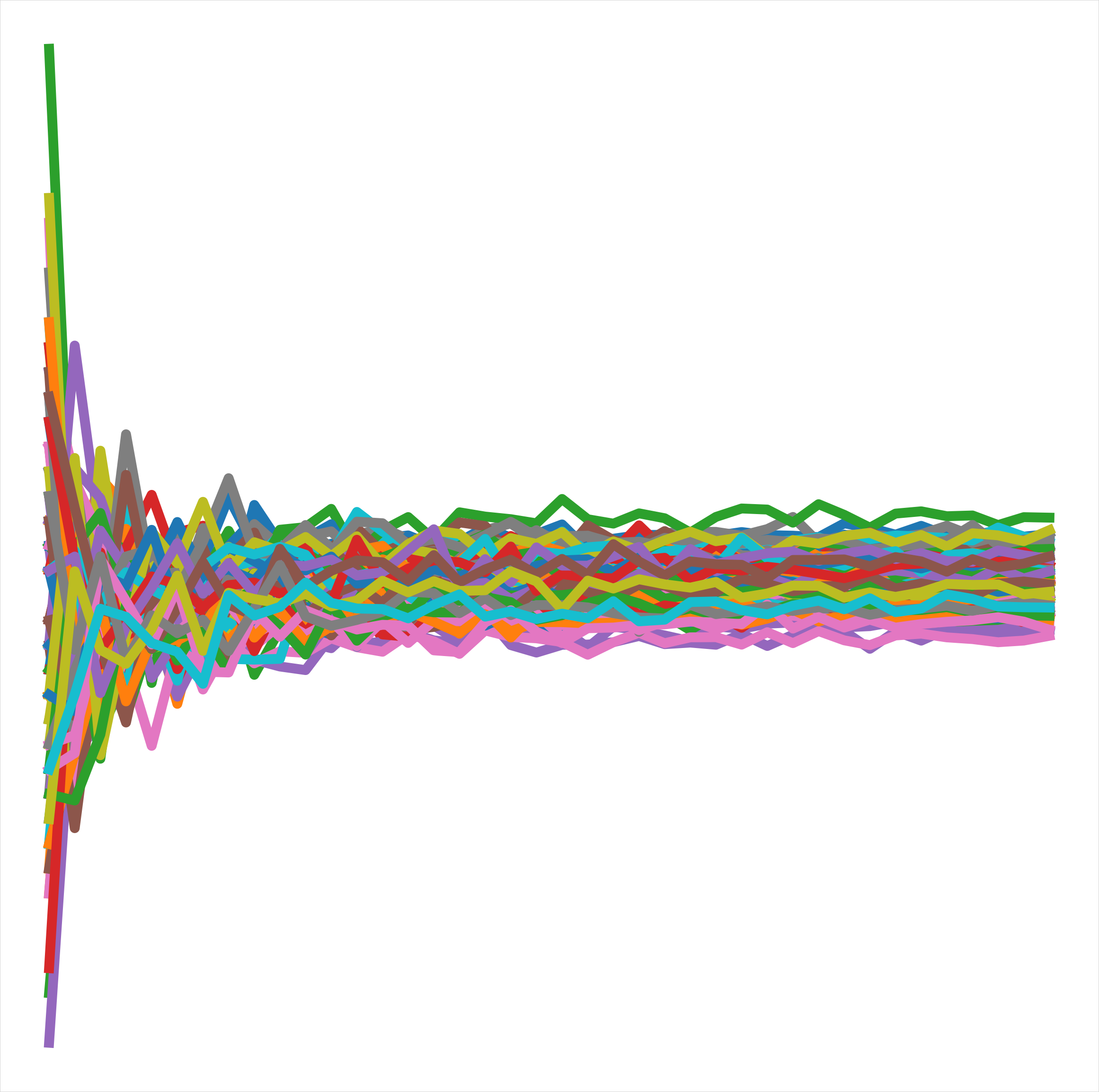};  
            \addlegendentry{y=0.33} 
            \end{axis}
        \end{tikzpicture}
        \caption{The unidirectional connection saturation of directed graphs of different scales gradually stabilizes}
        \label{fig:saturation}
    \end{subfigure}
    \hspace{0.02\textwidth}
    \begin{subfigure}[b]{0.4\textwidth}
        \centering
        \begin{tikzpicture}
            \begin{axis}[
                xlabel={n},   
                ylabel={y},   
                ylabel style={yshift=-1em},
                enlargelimits=false, 
                axis on top,         
                legend pos=north east, 
                width = 6cm,
                height = 6cm,
            ]
            {x}; 
            \addplot[thick,color={rgb:red,0.12;green,0.46;blue,0.7}] graphics[xmin=0, xmax=620, ymin=0, ymax=0.37] 
            {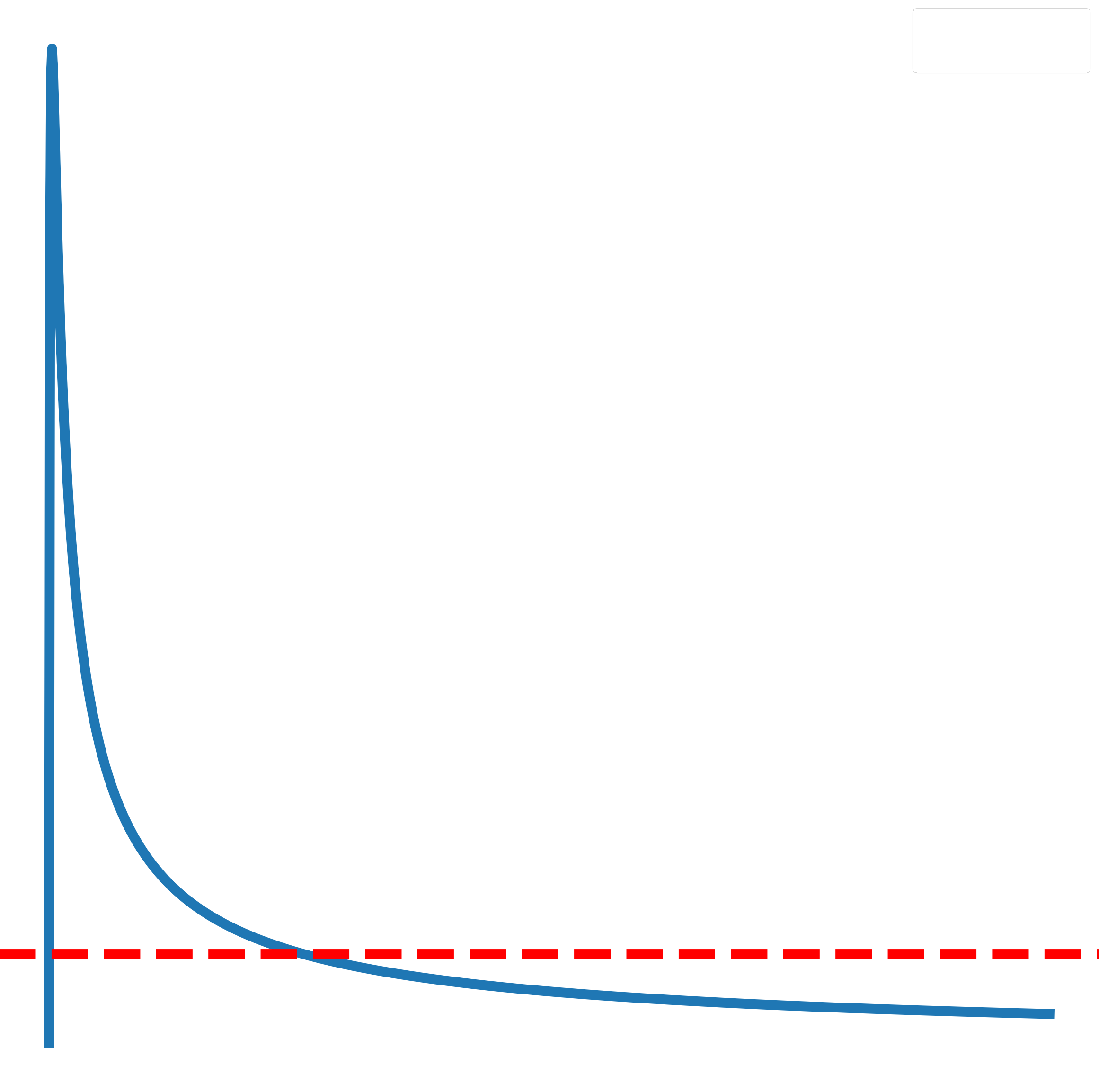};  
            \addplot[red,dashed,thick] graphics[xmin=0, xmax=620, ymin=0, ymax=0.37] 
            {pics/figure5};  
            \addlegendentry{y=1og(n)/n } 
            \addlegendentry{y=0.033} 
            \end{axis}
        \end{tikzpicture}
        \caption{Plot of $y=1og(n)/n$}
        \vspace{1em}
        \vspace{1em}
        \label{fig:logn/n}
    \end{subfigure}
    \hspace{0.02\textwidth}
    \begin{subfigure}[b]{0.4\textwidth}
        \centering
        \begin{tikzpicture}
            \begin{axis}[
                xlabel={Number of Relay Nodes},   
                ylabel={Ratio of Directed Graphs Have Loop},   
                ylabel style={yshift=-1em},
                enlargelimits=false, 
                axis on top,         
                legend pos=south east, 
                width = 6cm,
                height = 6cm,
            ]
            {x}; 
            \addplot[thick,color={rgb:red,0.12;green,0.46;blue,0.7}] graphics[xmin=80, xmax=200, ymin=0.2, ymax=1.04] 
            {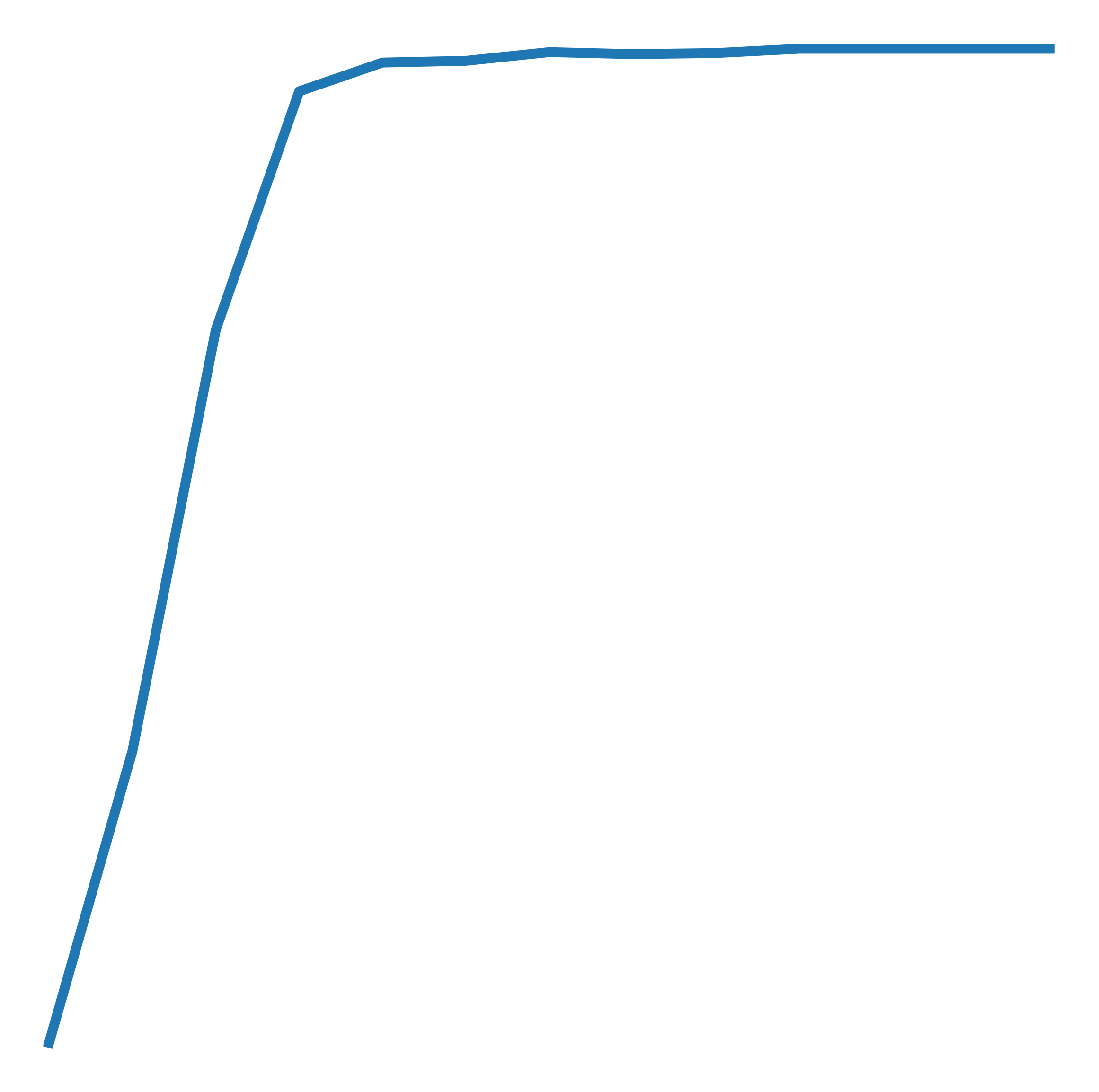};  
            \end{axis}
        \end{tikzpicture}
        \caption{Ratio of directed graphs with different numbers of relay nodes forming cycles within two hops.}
        \label{fig:rationOfDirected}
    \end{subfigure}
    \hspace{0.02\textwidth}
    \begin{subfigure}[b]{0.4\textwidth}
        \centering
        \begin{tikzpicture}
            \begin{axis}[
                xlabel={Scale of Number of Subgraph Nodes},   
                ylabel={Number of Hamiltonian Loops},   
                ylabel style={yshift=-1em},
                ymode=log,
                enlargelimits=false, 
                axis on top,         
                legend pos=north west, 
                width = 6cm,
                height = 6cm,
            ]
            \addplot[thick,color={rgb:red,0.12;green,0.46;blue,0.7}] graphics[xmin=0, xmax=84, ymin=1, ymax=1.2e16] 
            {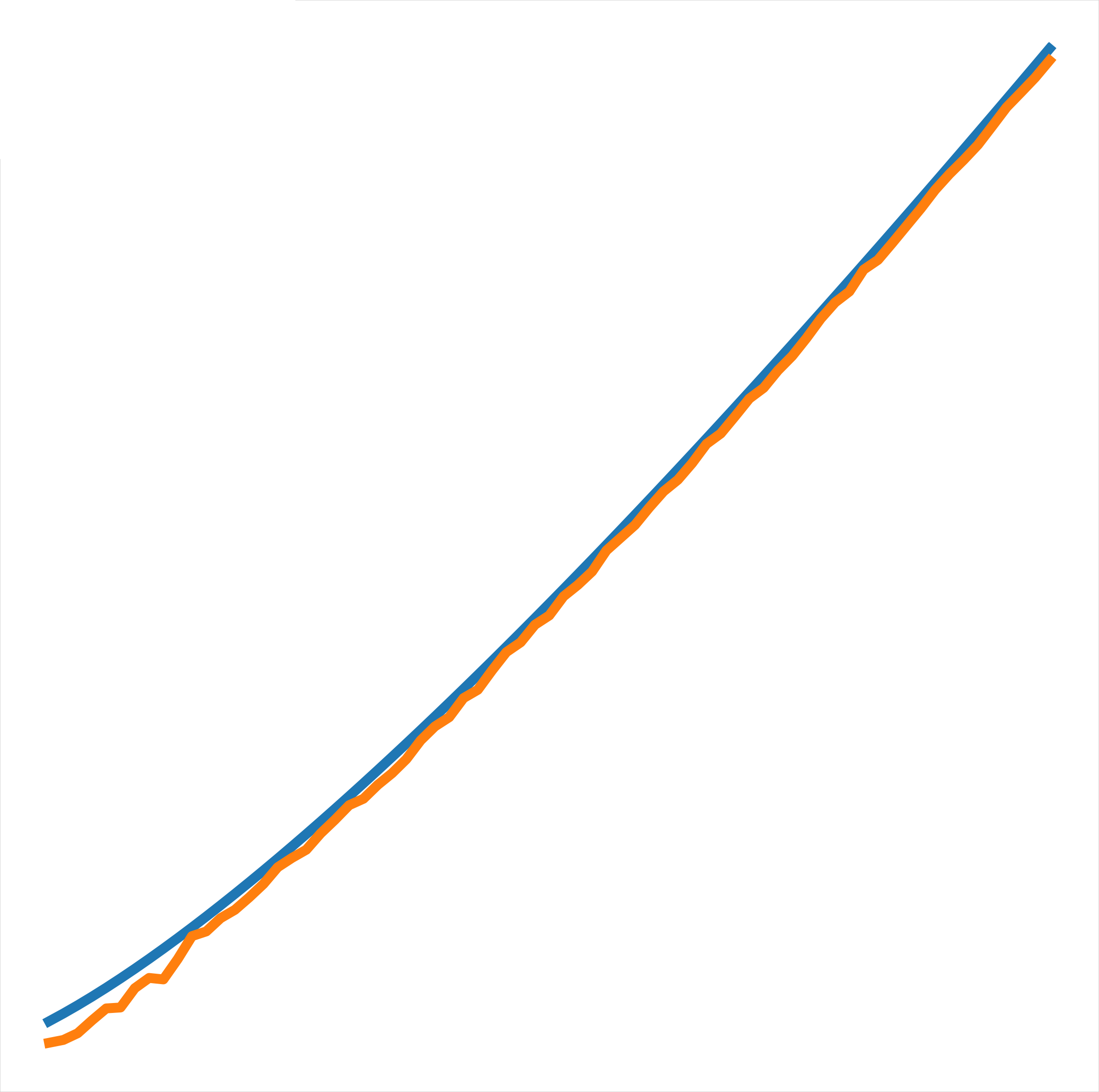};  
            \addplot[thick,color={rgb:red,0.99;green,0.49;blue,0.054}] graphics[xmin=0, xmax=84, ymin=1, ymax=1e17]
            {pics/newfigure345};  
            \addlegendentry{theoretical number} 
            \addlegendentry{experiment number} 
            \end{axis}
        \end{tikzpicture}
        \caption{Theoretical and experimental values of the number of Hamiltonian cycles in the node set}
        \label{fig:hamiAll}
    \end{subfigure}

    \begin{subfigure}[b]{0.4\textwidth}
        \centering
        \begin{tikzpicture}
            \begin{axis}[
                xlabel={Time(minutes)},   
                ylabel={Number of Connected Subgraphs},   
                ylabel style={yshift=-1em},
                ymode=log,
                enlargelimits=false, 
                axis on top,         
                legend pos=south east, 
                width = 6cm,
                height = 6cm,
            ]
            {x}; 
            \addplot[thick,color={rgb:red,0.12;green,0.46;blue,0.7}] graphics[xmin=0, xmax=960, ymin=1, ymax=1e8] 
            {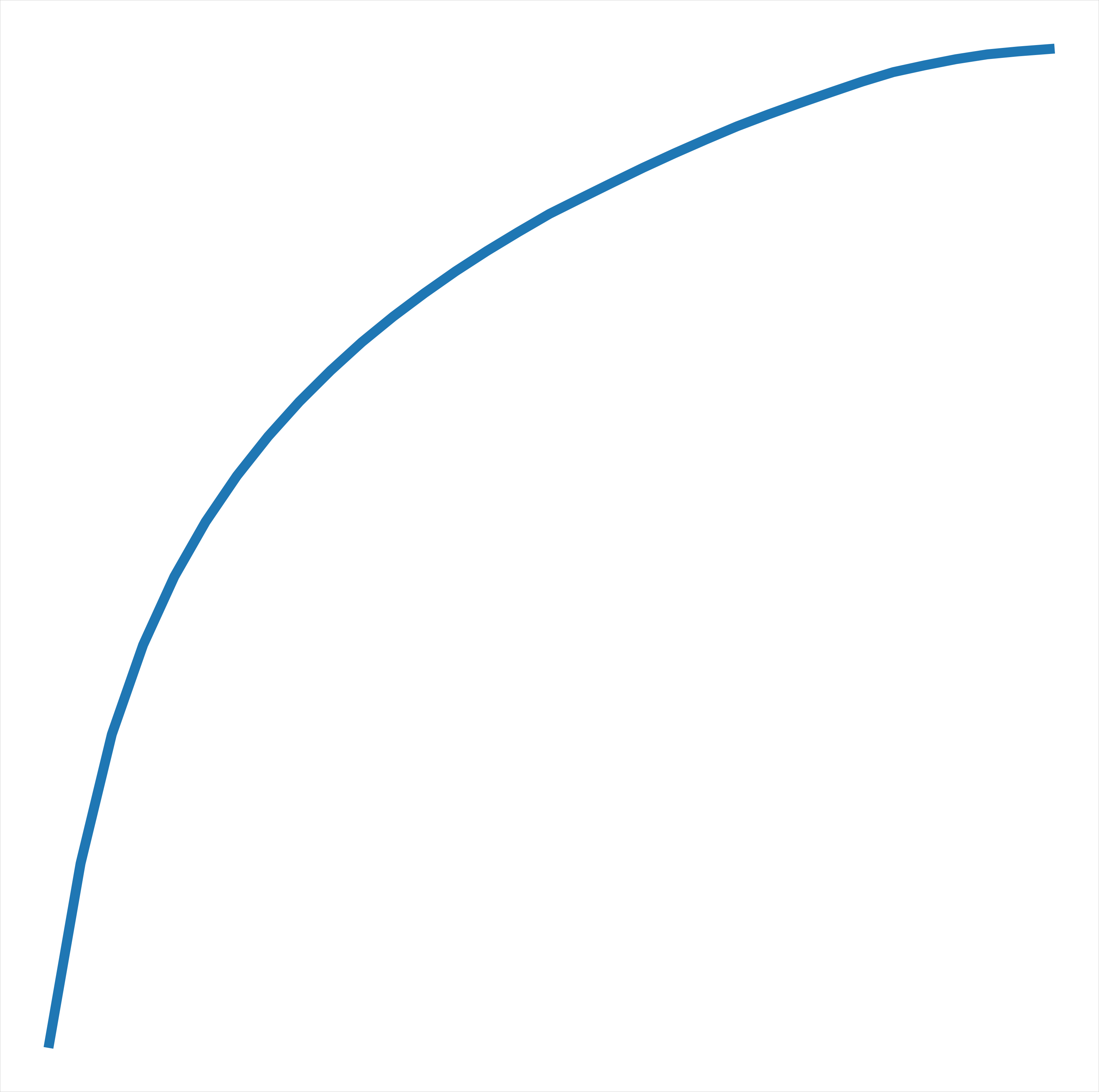};  
            \end{axis}
        \end{tikzpicture}
        \caption{Sampling experiment verification shows a huge number of available connected subgraphs.}
        \label{fig:weekconnected}
    \end{subfigure}
    \hspace{0.02\textwidth}
    \begin{subfigure}[b]{0.4\textwidth}
        \centering
        \begin{tikzpicture}
            \begin{axis}[
                xlabel={Similar Percentage(\%)},   
                ylabel={Cumulative Frequency},   
                ylabel style={yshift=-1em},
                enlargelimits=false, 
                axis on top,         
                legend pos=south east, 
                width = 6cm,
                height = 6cm,
            ]
            \addplot[red, dashed, thick] graphics[xmin=0, xmax=100, ymin=0, ymax=1.04] 
            {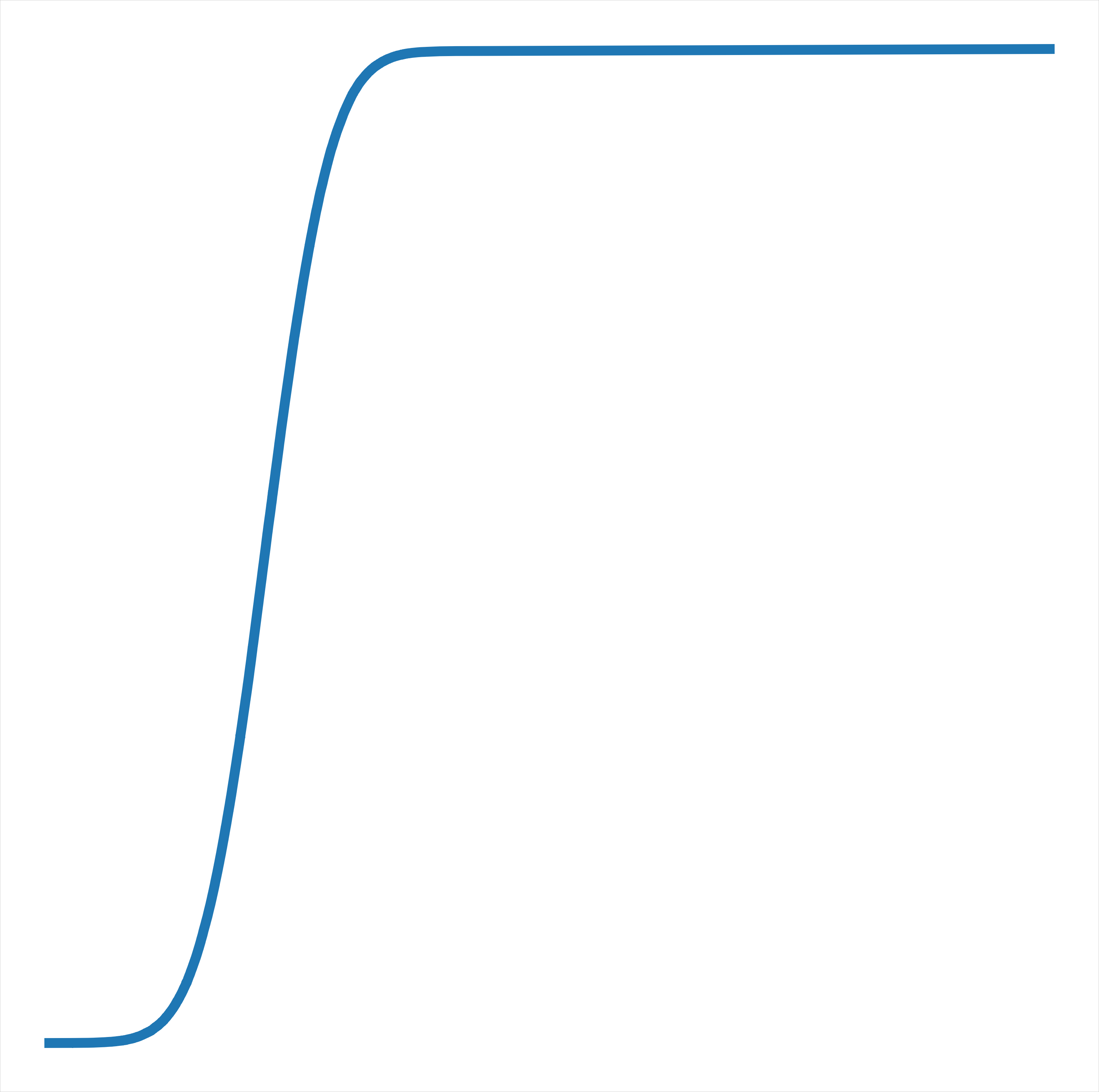};  
            \end{axis}
        \end{tikzpicture}
        \caption{Subgraph Similar Percentage Cumulative Distribution Function}
        \vspace{1em}
        \label{fig:cdf}
    \end{subfigure}
    \caption{}
\end{figure*}
So, do biological neural networks satisfy the constraints of this theorem? This paper uses unidirectional connection saturation to equivalently represent the connection probability $p$ between nodes in a random directed graph. Unidirectional connection saturation is defined as shown in \ref{eq:saturate}, where $M$ is the total number of edges in the directed graph, and $N$ is the number of nodes in the directed graph. Unidirectional connection saturation can macroscopically characterize the connectivity of the entire set of nodes. For a set of nodes, if its unidirectional connection saturation is high, then these nodes have better connectivity, and the likelihood of a connection path existing between any two nodes is greater.
\begin{equation}
    \left(S a t u r a t e\right)_{u n i d i r e c t i o n a l \textrm{ } c o n n e c t i o n} = \frac{M}{N \left(N - 1\right)}
    \label{eq:saturate}
\end{equation}
If the directed graph obtained according to \ref{eq:beconnected} (with parameters $k$=0.06 and $\lambda$=41) meets the Asaf-Ferber condition, the detailed experimental steps can be found in the section "SI Material and Methods: Experiment on the Relationship Between the Average Unidirectional Connection Saturation of Directed Graph Subgraphs and Subgraph Size". The conclusions are as follows.

The statistical results of 80,000 directed graphs of different scales generated based on \ref{eq:beconnected} are shown in \ref{fig:rationOfDirected}, with the x-axis representing the scale of the node set and the y-axis representing the average unidirectional connection saturation. From the figure, it can be observed that when the number of nodes exceeds approximately 50, the average unidirectional connection saturation of the node sets of different scales stabilizes around 0.033. Under the condition that parameters k and $\lambda$ remain unchanged, it approaches a constant. \ref{fig:logn/n} shows the graph of $y=\log{N}/{N}$. If we let $\log{N}/{N}<0.033$, then $N$>152.29. When $N$ exceeds 1, $\log{N}/{N}$ continuously decreases as $N$ increases, while 0.033 is a constant value. Therefore, theoretically, when the node set has 153 or more nodes, the directed graph will satisfy the Asaf-Ferber condition.

According to \ref{eq:saturate}, it is also possible to calculate how many Hamiltonian cycles exist when the node set has 153 or more nodes. Since finding Hamiltonian cycles in directed graphs is an NP-complete problem, the time complexity of traversal algorithms is very high, approximately $o\left(N!\right)$, making it infeasible to obtain results in a reasonable time for larger $N$. To address this issue, this paper uses a computer to enumerate Hamiltonian cycles in directed graphs when the number of nodes is small. For larger node counts, the Monte Carlo method is used to estimate the number of Hamiltonian cycles present in the directed graph. The specific experimental strategy can be found in the section "SI Material and Methods: Counting Hamiltonian Cycles in Directed Graphs".

The overall results are shown in \ref{fig:hamiAll} where the theoretical values closely match the measured values. This experiment reveals that the magnitude of Hamiltonian cycles contained in the directed graph subgraphs generated by the biological neuron connection probability model, as well as the trend of variation in the number of cycles contained in subgraphs with different numbers of nodes, is similar to that predicted by the Asaf-Ferber formula. Thus, the directed graph model proposed in this paper theoretically possesses a substantial number of Hamiltonian cycles in its subgraphs, providing significant convenience for constructing connected subgraphs.

\subsection{The formation cost of connected subgraphs is relatively low}
In the subgraph of a directed graph, cycles can enhance the robustness of connections, forming closed-loop structures that ensure all relevant nodes are interrelated, preserving the possibility of being awakened and awakening others. The previous discussion has demonstrated that the subgraphs of directed graph proposed in this paper, which meet biological constraints, theoretically possess a substantial number of Hamiltonian cycles. However, the detail structure organization of different individuals' brains is not identical, particularly at the lower levels, such as the physical morphology of the neural circuit, yet they can still retain the same memories. This necessitates that directed graphs with different structural details can connect semantically similar nodes (such as neurons representing colors in the visual cortex or neurons representing audio in the auditory cortex) through different physical pathways. 

At the same time, connectivity incurs costs, including hardware costs related to length and signal transmission time costs, which means that the connecting paths should not be excessively long. This can be represented using the concept of "hops" in graph theory. This paper experimentally verifies (see SI Material and Methods) the likelihood of initially activated nodes forming cycles within two hops, with the experimental results shown in \ref{fig:rationOfDirected}. It can be observed that, under the current node connection probability, as long as there is a sufficient number of relay node, those initially activated nodes can be rapidly connected through local edges, thereby forming robust connected subgraphs. A more rigorous mathematical conclusion is that, under the premise of the connection probability defined in \ref{eq:beconnected}, given $M$ uniformly distributed nodes, the mathematical expectation of the distance scale between nearest neighbor nodes is $\sqrt{s}$.Therefore, the probability that this subset of nodes fails to connect and forms two independent branches (where $M=M_1+M_2$) is: 
\begin{equation}
    \prod_{x=1}^{M_1} (1-P(be\_connected)_{d_x=J\sqrt{s} }) ^{M_2},J\in\{1,2,3,...,M\} 
\end{equation}
This is a product of exponential functions, and when $M\geq 20$, the probability becomes very low. Furthermore, if the connection failure results in the formation of three isolated branches (where $M=M_1+M_2+M_3$), the probability is:
\begin{equation}
    \begin{split}
        \prod_{x=1}^{M_1} (1-P(be\_connected)_{d_x=J_1\sqrt{s} }) ^{M_2}\\
        \times \prod_{x=1}^{M_1} (1-P(be\_connected)_{d_x=J_2\sqrt{s} }) ^{M_3}\\
        \times \prod_{x=1}^{M_2} (1-P(be\_connected)_{d_x=J_3\sqrt{s} }) ^{M_3},\\
        J_1,J_2,J_3\in\{1,2,3,...,M\}
    \end{split}
\end{equation}
This value is also very small (see supplementary information for details). Similarly, the probability of forming more isolated branches can be derived. Therefore, the probability of successfully connecting a set of nodes with a certain size is guaranteed.

\subsection{The potential available capacity is enormous}
A directed graph with a substantial number of nodes, when considering the combinations of nodes, will have an extremely large number of subgraphs. Based on the aforementioned connection probability between nodes, the likelihood of these subgraphs being connected is high. This implies that a directed graph can have an immense theoretical capacity. For example, in a directed graph formed by 540 nodes, if 20\% of the nodes are selected to form connected subgraphs, the possible number of combinations is $C_{540}^{108}\approx  1.576\times10^{85}$.If an event represented by 108 nodes is described by a statement composed of six words, the theoretical capacity limit would be equivalent to $1.17\times10^{77}$ volumes of the Encyclopedia Britannica (digitized with text and images, approximately 6$GB$, assuming the average length of an English word is eight letters).

Furthermore, based on the discussion in C, it is theoretically known that when the node set contains 153 or more nodes, the directed graph satisfies the Asaf-Ferber condition, meaning it contains at least one Hamiltonian cycle. Although 153 nodes can theoretically produce approximately $2\times10^{269}$ different cycles, even in a directed graph with a unidirectional connection saturation of 3\%, according to Asaf-Ferber's formula, there are still about $2\times10^{36}$ Hamiltonian cycles, which is still an enormous number. Hamiltonian circuits impose overly strict requirements on connectivity, but in fact, weakly connected graphs can also serve as memory carriers. It is impossible for us to explore such a vast space within a limited time, and sampling is the only feasible experimental approach. Specifically, we perform random sampling of the edges between these 153 nodes and determine whether they can form weak connectivity. In a continuous 960-minute sampling experiment, we accumulated over $10^7$ connected subgraphs, as shown in \ref{fig:weekconnected}. Notably, the number of connected subgraphs continues to grow, and the slope does not show signs of decreasing, indicating that this directed graph as a whole possesses immense potential capacity.

Considering distinguishability and the sparse usage of the combinatorial space, two subgraphs should maintain sufficient differentiation. If subgraph $A$ and subgraph $B$ have edge sets ${Edge}_A$ and ${Edge}_B$, the similarity between two subgraphs is defined by the overlap ratio of their edge sets:
\begin{equation}
    S i m i l a r \left(\right. A , B \left.\right) = \left|\frac{\left(E d g e\right)_{A} \cap \left(E d g e\right)_{B}}{\left(E d g e\right)_{A} \cup \left(E d g e\right)_{B}}\right|
\end{equation}
Based on the aforementioned directed graph, if the similarity between any two subgraphs is required to be no greater than 40\%, the experimental results, as shown in \ref{fig:cdf}, indicate that approximately 55\% of the subgraphs can meet this requirement. Although this significantly reduces the theoretical capacity, it still provides a considerable amount of practical usable capacity.

\subsection{Incremental Achievability}
A directed graph can be seen as a collection of many stacked connected subgraphs, each representing a specific event through a particular topological structure \cite{Wei2002homogenous,Wei2004two-layer}, This is fundamentally different from current machine learning methods (like the 2024 Nobel Prize in Physics and Chemistry), which bundle a large number of highly homogeneous and uniform samples, perform integrated learning, and use weights to represent them, essentially functioning as function approximation. The former allows for easy separation of the original samples, while the latter is difficult to separate, and the former provides better interpretability. The content of memory is incrementally added, making directed graphs more suitable for this "record-as-it-comes" method. Learning dispersed across the internal nodes can consolidate the topological structure of individual connected subgraphs\cite{wei2023autonomous,wei2024modeling,wei2023storage}.

\section{Methods}
\subsection{Mathematical Model of Connection Probability Between Cortical Neurons}
Based on the synaptic density field, the number of synapses between two neurons was derived, and further, the expression for the connection probability between two neurons was obtained, as shown in equation \ref{eq:beconnected} (for details, see SI Material and Methods: Neuron Connection Probability Modeling).
\subsection{Modeling Directed Graphs that Meet Biological Constraints}
Using the neuron probability model proposed in this paper as the basis for generating directed graphs, connectivity performance was tested with metrics such as reachability, average path length, and clustering coefficient for directed graphs with different numbers of nodes. It was found that directed graphs with 500 nodes exhibited superior connectivity performance. Therefore, a directed graph with 500 nodes was selected for subsequent experiments (for details, see SI Material and Methods: Modeling Directed Graphs that Meet Biological Constraints).
\subsection{Experiment on the Relationship Between the Average Unidirectional Connection Saturation of Directed Graph Subgraphs and Subgraph Size}
Setting the directed graph parameters to $k=0.06$ and $\lambda=41$, the experiment was designed to statistically analyze the average unidirectional connection saturation of node subsets of different scales within the directed graph. The experiment confirmed that when the size of the node subset reaches a certain scale, the average connection saturation stabilizes around a fixed value (for details, see SI Material and Methods: Experiment on the Relationship Between the Average Unidirectional Connection Saturation of Directed Graph Subgraphs and Subgraph Size).
\subsection{Counting Hamiltonian Cycles in Directed Graphs}
When the number of nodes is small, computer enumeration was used to count the number of Hamiltonian cycles in the directed graph. For larger numbers of nodes, the Monte Carlo method was used to estimate the number of Hamiltonian cycles. The experiment verified that both methods produced approximately consistent results, with the experimental results shown in \ref{fig:hamiAll} (for details, see SI Material and Methods: Counting Hamiltonian Cycles in Directed Graphs).

\section*{Acknowledgments}
This work was supported by the NSFC Project (Project No. 61771146).

\bibliographystyle{apalike} 
\bibliography{references}

\appendix  
\section*{Supplementary Materials}

\subsection*{Neuron Connection Probability Modeling}

\begin{figure*}[thbp]
 \centering
    \begin{subfigure}[b]{0.45\textwidth}
        \centering
        \includegraphics[width=\textwidth]{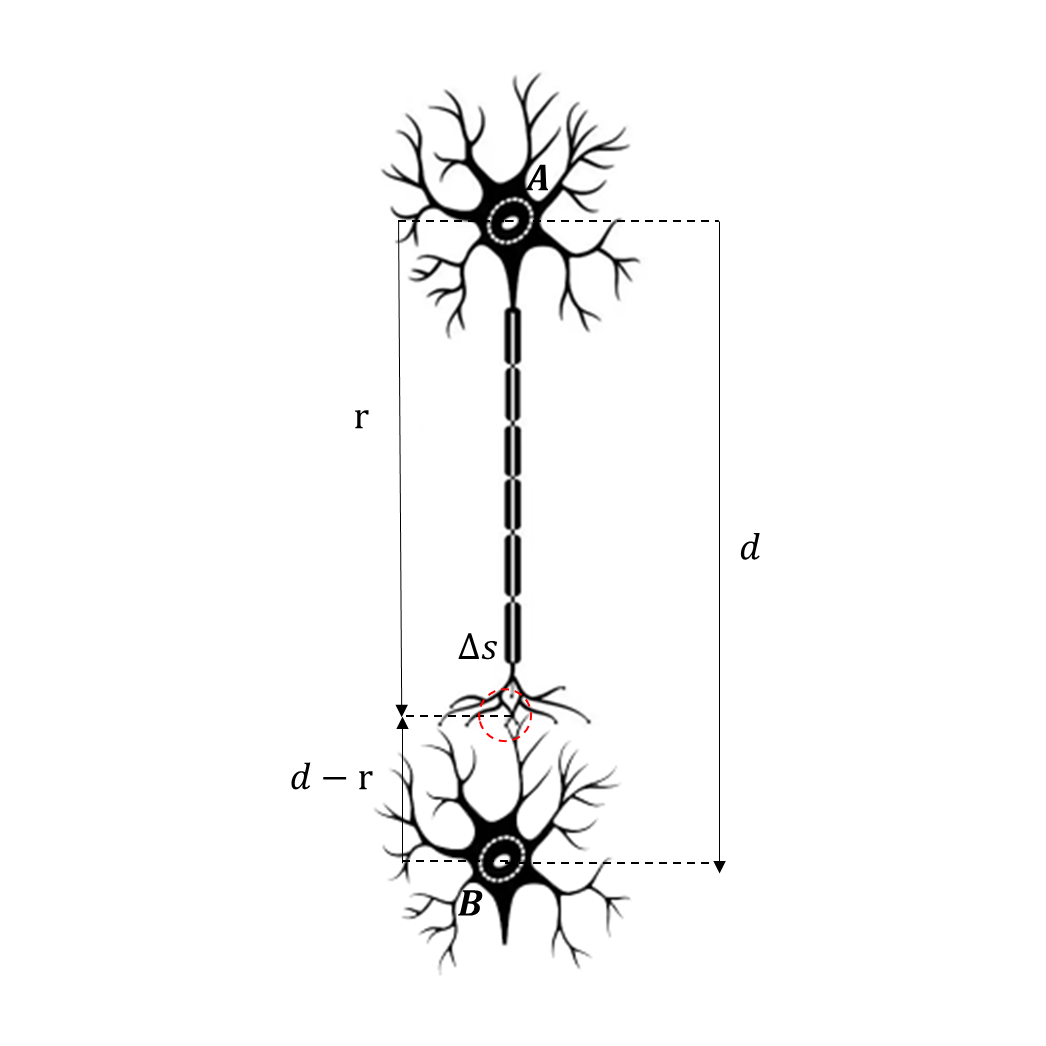}
        \caption{Neighborhoods of Two Neurons}
        \label{fig:neighbor}
    \end{subfigure}
    \hspace{0.02\textwidth}
    \begin{subfigure}[b]{0.45\textwidth}
        \centering
        \includegraphics[width=\textwidth]{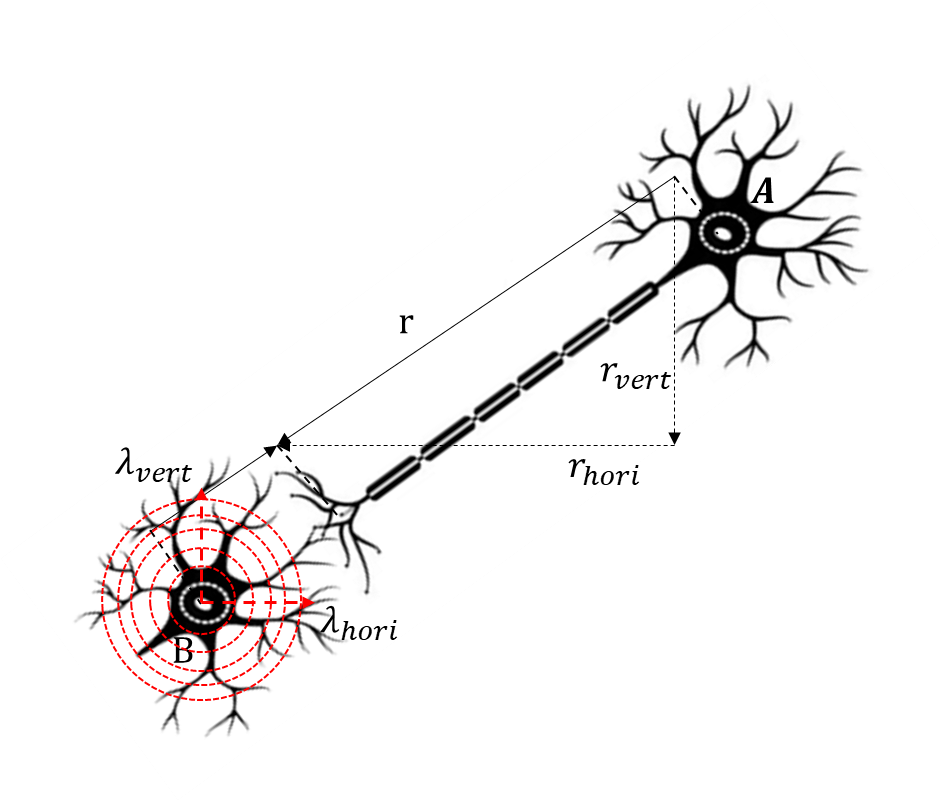}
        \caption{Diagram of Synaptic Field Density}
        \label{fig:density}
    \end{subfigure}
    \caption{Modeling of Neuronal Connections}
\end{figure*}
\begin{figure*}[thbp]
    \centering
    \begin{subfigure}[b]{0.3\textwidth}
        \centering
        \begin{tikzpicture}
            \begin{axis}[
                xlabel={Number of Nodes},   
                ylabel={Reachability},   
                ylabel style={yshift=-1em},=
                enlargelimits=false, 
                axis on top,         
                legend pos=south east, 
                width = 6cm,
                height = 6cm,
            ]
            \addplot[thick,color={rgb:red,0.12;green,0.46;blue,0.7}] graphics[xmin=60, xmax=860, ymin=0.35, ymax=1.05] 
            {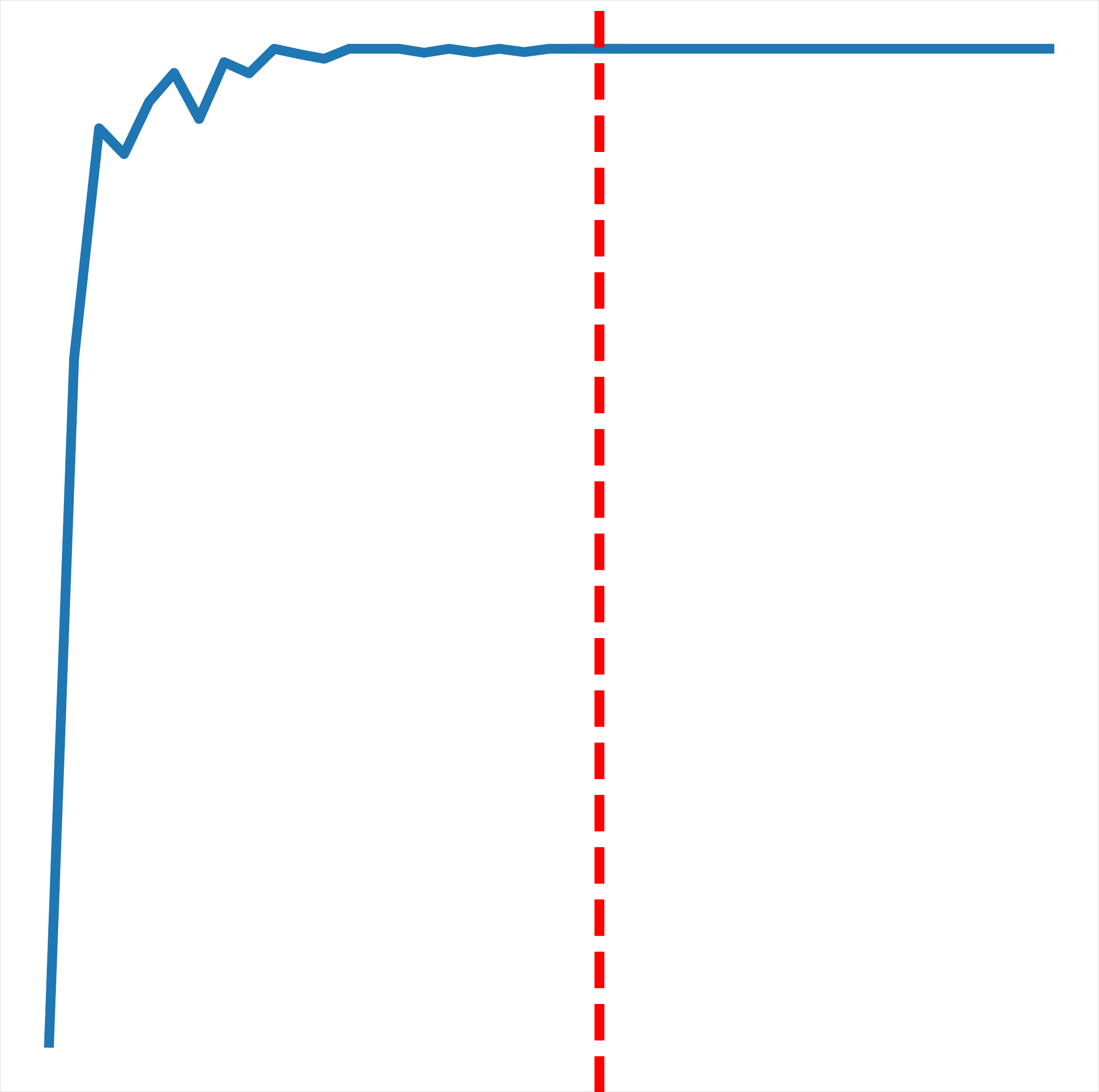};  
            \end{axis}
        \end{tikzpicture}
        \caption{Statistical results of reachability for directed graphs of different node sizes}
        \label{fig:reachability}
    \end{subfigure}
    \hspace{0.02\textwidth}
    \begin{subfigure}[b]{0.3\textwidth}
        \centering
        \begin{tikzpicture}
            \begin{axis}[
                xlabel={Number of Nodes},   
                ylabel={Average Path Length},   
                ylabel style={yshift=-1em},=
                enlargelimits=false, 
                axis on top,         
                legend pos=south east, 
                width = 6cm,
                height = 6cm,
            ]
            \addplot[thick,color={rgb:red,0.12;green,0.46;blue,0.7}] graphics[xmin=60, xmax=860, ymin=2, ymax=5.8] 
            {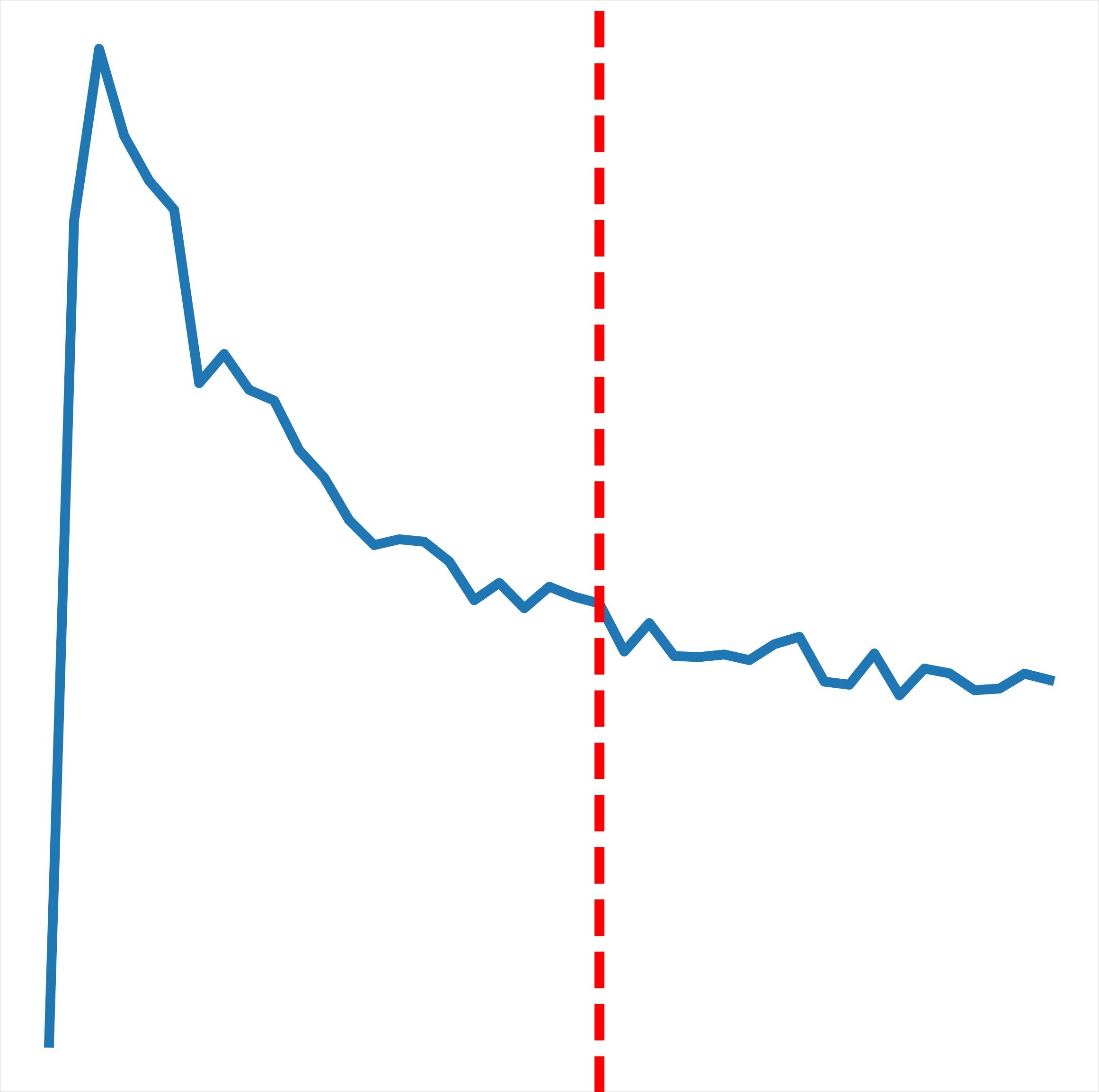};  
            \end{axis}
        \end{tikzpicture}
        \caption{Statistical results of average path length for directed graphs of different node sizes}
        \label{fig:avgPath}
    \end{subfigure}
    \hspace{0.02\textwidth}
    \begin{subfigure}[b]{0.3\textwidth}
        \centering
        \begin{tikzpicture}
            \begin{axis}[
                xlabel={Number of Nodes},   
                ylabel={Clustering Coefficient},   
                ylabel style={yshift=-1em},=
                enlargelimits=false, 
                axis on top,         
                legend pos=south east, 
                width = 6cm,
                height = 6cm,
            ]
            \addplot[thick,color={rgb:red,0.12;green,0.46;blue,0.7}] graphics[xmin=60, xmax=860, ymin=0.215, ymax=0.35] 
            {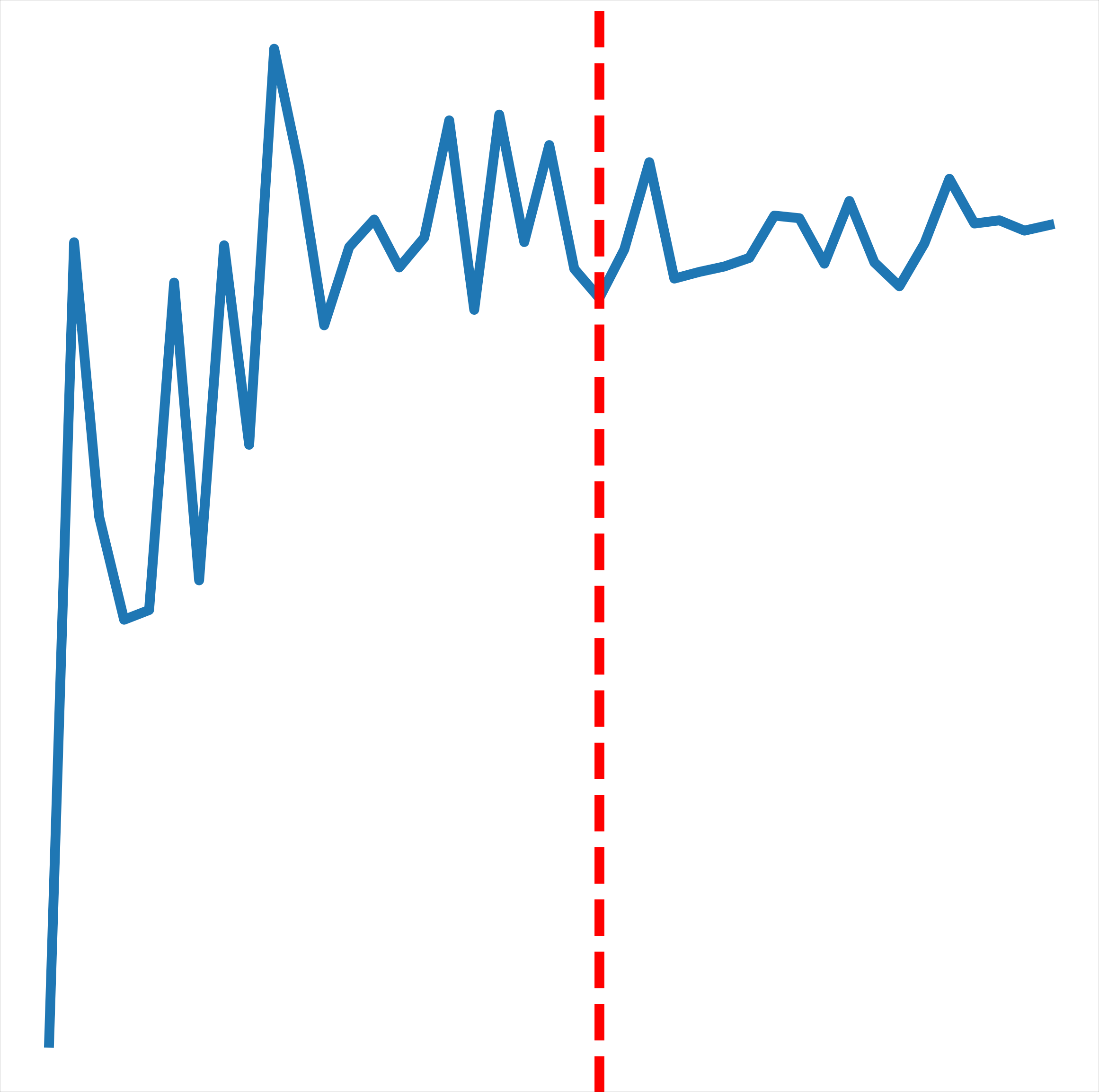};  
            \end{axis}
        \end{tikzpicture}
        \caption{Statistical results of clustering coefficient for directed graphs of different node sizes}
        \label{fig:clustercoefficient}
    \end{subfigure}
    \caption{}
\end{figure*}
Neurons connect to each other through synapses. If two neurons, A and B, have a sufficient number of synapses within a sufficiently small neighborhood, the probability of a connection between A and B increases. To determine the probability of connection between any two neurons, whether they are near or far, this paper treats the formation of synapses at a particular spatial location as an independent event. if two neurons have enough synapses within a sufficiently small neighborhood, then it is possible for these two neurons to form synaptic connections within that neighborhood\cite{tecuatl2021comprehensive}.

Since neurons are very small and can be approximated as thin layers with dendritic spines compressed into this layer, two-dimensional coordinates are sufficient to represent their positions. As shown in \ref{fig:neighbor}, the distance between a presynaptic neuron $A$ and a postsynaptic neuron $B$ is denoted as $d$. Suppose neuron $A$ is placed at the origin, and its axon has a length $r$. Neuron $B$ is positioned at a distance $d$. They may have multiple neighborhoods close enough to form connections, and the location of each neighborhood depends on the values of $d$ and $r$.
  
To represent the potential number of synapses in a certain neighborhood, the concept of synaptic field density is introduced \cite{2005A}. Synaptic field density refers to the density of synapses within a specific area, that is, the number of synapses per unit area. Let $\eta\left(r\right)$ be the number of synapses on the cell membrane of area $\mathrm{\Delta s}$ at distance $r$, and $\rho(r)$ be the synaptic field density. Then:
\begin{equation}
    \rho\left(r\right)=\ \frac{\eta\left(r\right)}{\Delta s}	
    \label{eq:density}
\end{equation}

Let $\xi$ be the proportion of upstream and downstream synapses that may connect to each other within the neighborhood $\mathrm{\Delta s}$. If $\mathrm{\Delta s}$ is sufficiently small, then the number of synapses that upstream and downstream neurons can form at this location can be simplified to either 0 or 1. Let $\eta\left(r\right)$ and $\eta\left(d-r\right)$ represent the number of synapses of neurons $A$ and $B$ in the neighborhood at $r$. Then, the density of the number of synaptic connections formed in $\Delta s$ can be expressed as:
\begin{equation}
\sigma \left(r , d\right) = \xi \textrm{ } \frac{\eta \left(r\right) \cdot \eta \left(r - d\right)}{\Delta s}
    \label{eq:densityofrd}
\end{equation}
Here, let $\Delta s\cdot\ \xi$ be considered as a constant $A$. Then, from \ref{eq:densityofrd}., we have:
\begin{equation}
    \begin{split}
    \sigma \left(r , d\right) &= \xi \cdot \Delta s \textrm{ } \frac{\eta \left(r\right) \cdot \eta \left(r - d\right)}{\Delta s \cdot \Delta s}\\
    &= A \frac{\eta \left(r\right)}{\Delta s} \textrm{ } \frac{\eta \left(r - d\right)}{\Delta s}\\
    &= \mathsf{A} \cdot \rho \left(r\right) \cdot \rho \left(d - r\right)\\
    \end{split}
\end{equation}
The variable $r$ represents the length covered by the axon and its terminals of the upstream neuron. The upstream neuron may have multiple axonal terminals that are in close proximity to the dendritic spines of multiple downstream neurons, thus forming connections within several small neighborhoods. Therefore, by integrating over $r$, we can obtain a function that represents the number of synaptic connections between upstream and downstream neurons:
\begin{equation}
    N \left(d\right) = \mathsf{A} \int \rho \left(r\right) \cdot \rho \left(d - r\right) d r
    \label{eq:N(d)}
\end{equation}
Therefore, to further calculate $N\left(d\right)$, we also need to obtain the synaptic densities $\rho(r)$ and $\rho(d-r)$ at two locations. These can be approximated as follows: First, the cerebral cortex is a layered structure, and the physical properties of the cortex are isotropic in the horizontal dimension; thus, synaptic density can also be considered isotropic. Therefore, the synaptic density field can be modeled as an ellipsoidal field, as shown by the red dashed area in \ref{fig:density}. Second, research has found that the number of synapses decreases exponentially with respect to the longitudinal and transverse axes of the ellipsoid \cite{1955Sholl}. Thus, we can set
\begin{equation}
    \rho \left(r\right) = \left(\rho\right)_{0} e^{- \sqrt{\frac{r_{h o r i}^{2}}{\lambda_{h o r i}^{2}} + \frac{r_{v e r t}^{2}}{\lambda_{v e r t}^{2}}}}
\end{equation}
where$\ \rho_0$ is the synaptic density at the center of the ellipsoid, $\vec{r_{hori}}$ and $\vec{r_{vert}}$ are the horizontal and vertical components of $\vec{r}$, respectively, with $\vec{r}=\vec{r_{hori}}+\vec{r_{vert}}$. The decay rate is denoted by $\vec{\lambda}$, where $\vec{\lambda}=\vec{\lambda_{hori}}+\vec{\lambda_{vert}}$. To simplify the calculations, we can assume that the decay rates of synaptic density in the horizontal and vertical directions are the same, so $\lambda_{hori}^2=\lambda_{vert}^2$, which gives $\left|\vec{\lambda_{hori}}\right|=\left|\vec{\lambda_{vert}}\right|={\sqrt2}/{2}\vec{\lambda}$. Then,
\begin{equation}
    \begin{split}
    \rho \left(r\right) &= \left(\rho\right)_{0} e^{- \sqrt{\frac{\left(r_{h o r i}^{2} + r_{v e r t}^{2}\right)}{\lambda_{v e r t}^{2}}}}\\
    &= \left(\rho\right)_{0} e^{- \sqrt{\frac{\left(r_{h o r i}^{2} + r_{v e r t}^{2}\right)}{\frac{1}{2} \lambda^{2}}}}\\
    &= \left(\rho\right)_{0} e^{- \sqrt{\frac{2 r^{2}}{\lambda^{2}}}}\\
    \end{split}
\end{equation}
Thus, substituting into equation \ref{eq:N(d)}, have:
\begin{equation}
    \begin{split}
        N \left(d\right) &= \mathsf{A} \int \rho \left(r\right) \cdot \rho \left(d - r\right) d r\\
        &= \mathsf{A} \int \rho_{0} e^{- \sqrt{\frac{\left(2 r\right)^{2}}{\lambda^{2}}}} \cdot \rho_{0} e^{- \sqrt{\frac{\left(2 \left(d - r\right)\right)^{2}}{\lambda^{2}}}} d r\\
        &= \mathsf{A} \left(\rho_{0}\right)^{2} \int e^{- \frac{\sqrt{2 \left(r^{2} + \left(d - r\right)^{2}\right)}}{\lambda}} d r \textrm{ }\\
    \end{split}
    \label{eq:putinN(d)}
\end{equation}

Since equation \ref{eq:putinN(d)} does not yield an analytical solution, we can only obtain a numerical solution for practical use, specifically by calculating the value of $N\left(d\right)$ for $r\in[0,d]$ This can be expressed as a definite integral as shown in equation\ref{eq:definiteIntegralN(d)}:
\begin{equation}
    N \left(d\right) = \mathsf{A} \left(\rho_{0}\right)^{2} \int_{0}^{d} e^{- \frac{\sqrt{2 \left(\right. r^{2} + \left(d - r\right)^{2} \left.\right)}}{\lambda}} d r
    \label{eq:definiteIntegralN(d)}
\end{equation}
Using the first-order Euler approximation, the above definite integral can be transformed into the sum of the areas of multiple small rectangles, where $\Delta r$ is taken to be a very small value:

\begin{equation}
    N \left(d\right) = \mathsf{A} \left(\rho_{0}\right)^{2} \sum e^{- \frac{\sqrt{2 \left(\right. r^{2} + \left(d - r\right)^{2} \left.\right)}}{\lambda}} \Delta r \textrm{ }
\end{equation}
Finally, we need to convert the possible number of synapses between two neurons into their connection probability. Research has shown that a Poisson distribution can be used to fit the relationship between the possible number of synapses and the connection probability between neurons \cite{kersen2022connectivity}, that is:
\begin{equation}
    P \left(x = n\right) = \frac{\left(N \left(d\right)\right)^{n}}{n !} e^{- N \left(d\right)}
\end{equation}

where $x$ refers to the actual number of connections between neurons. Two neurons may have multiple synaptic connections, for example, the axon terminals of the upstream neuron may simultaneously form connections at various locations of the downstream neuron, such as the soma, the distal or proximal dendrites, or the axon hillock. In this paper, when modeling directed graphs, these synaptic connections are combined into a single edge, meaning that the connection probability between the two neurons is the probability that there is at least one synaptic connection between them, i.e., the probability is greater than zero:

\begin{equation}
    \begin{split}
        P \left(be\_connected\right) &= P \left(x \neq 0\right) = 1 - P \left(x = 0\right)\\
        &= 1 - \frac{\left(N \left(d\right)\right)^{0}}{0 !} e^{- N \left(d\right)}\\
        &= 1 - e^{- N \left(d\right)}\\
        &= 1 - e^{- \mathsf{A} \left(\rho_{0}\right)^{2} \sum e^{- \frac{\sqrt{2 \left(r^{2} + \left(d - r\right)^{2}\right)}}{\lambda}} \Delta r \textrm{ } \textrm{ }}\\
    \end{split}
\end{equation}

Let $k = A\left(\rho_{0}\right)^{2}$,
\begin{equation}
    P \left(b e\_{c o n n e c t e d}\right) = 1 - e^{- k \sum e^{- \frac{\sqrt{2 \left(r^{2} + \left(d - r\right)^{2}\right)}}{\lambda}} \Delta r \textrm{ }}
    \label{eq:beconnected}
\end{equation}
Through the above derivation, the connection probability between neurons, influenced by factors such as the distance $d$ between neurons, and the connection parameters $k$ and $\lambda$, has been analytically characterized. Next, we can utilize this formula to further construct directed graphs that authentically reflect the findings of neurobiological experiments.
\subsection*{Modeling Directed Graphs that Meet Biological Constraints.}
Using the neuron connection probability model proposed above as the basis for generating directed graphs, an enumeration experiment revealed that when the parameters are set to $k=0.06$ and $\lambda=41$, the neuron connection probability fits the data from Budd et al. \cite{budd2001local}. Therefore, by setting the directed graph parameters to $k=0.06$ and $\lambda=41$, a directed graph model that meets biological constraints can be established. To test whether the generated directed graphs exhibit statistically significant connectivity, 400 directed graphs with different topological structures were generated within a $1000 \times 1000 $spatial range, with node counts ranging from 100 to 900 in increments of 20. The connectivity performance was measured using three metrics: reachability, average path length, and clustering coefficient.

Reachability: A node $B$ is considered reachable from node $A$ if there exists a directed path that allows traversal from $A$ to $B$. This paper assesses the reachability of the directed graph by calculating the percentage of nodes that can be reached on average.

Average Path Length: The average path length refers to the mean shortest path length between any two nodes in the graph. This paper evaluates the average path length of the directed graph by computing the average of the shortest path lengths between all pairs of nodes.

Clustering Coefficient: For any given node $v_i$, the local clustering coefficient is defined as: $C\left(v_i\right)={e_i}/{k_i\ast{(k}_i-1)}$where $e_i$ is the number of edges between the node and its neighbors, and $k_i$ is the number of neighbors of the node. This paper uses the average local clustering coefficient of all nodes in the directed graph as an evaluation metric for the clustering degree of the graph.

Figure \ref{fig:reachability} shows the reachability statistics of directed graphs with different node scales. As the node distribution becomes denser, the reachability of the directed graphs also increases. Once the number of nodes reaches a scale of 500, the graph even becomes fully connected. 

Figure \ref{fig:avgPath} presents the average path length statistics of directed graphs with varying node scales. In scenarios with sparse node distribution, many nodes in the directed graph remain unconnected, resulting in a large number of nodes being unreachable. Only a few pairs of nearby nodes are connected, leading to a small average path length. As the scale of the nodes increases and the reachability of the directed graph improves, the average path length initially increases and then decreases, with the rate of decrease gradually slowing down, ultimately stabilizing around 3.6. 

Figure \ref{fig:clustercoefficient} illustrates the clustering coefficient statistics of directed graphs with different node scales. When the node scale is small, the connectivity of the directed graph is poor, resulting in a low clustering coefficient. As the node scale increases, the clustering coefficient of the directed graph gradually stabilizes around 0.32. In the anatomical experiments conducted by Kramer et al. based on engram cells, results similar to theoretical predictions were observed \cite{kramer2023spatial}.

From the experiments evaluating the metrics of directed graphs with different node sizes, it can be seen that a directed graph with 500 nodes exhibits high reachability, short average path length, and high clustering coefficient. This aligns with data from the cat visual cortex (area 17) in layer 4, where the average density of neurons is approximately 54,000 per cubic millimeter. Taking 1\% of that, in a thin layer measuring 1000 $\mu \text{m} \times 1000 \, \mu \text{m} \times 10 \, \mu \text{m}$, there are about 540 cells. Since 10 micrometers is very thin and almost equivalent to a plane, it can be assumed that these 540 cells are distributed across this two-dimensional plane. The experiment also shows that a directed graph with 540 nodes possesses high reachability, short average path length, and high clustering coefficient, which facilitates information exchange between neurons. This density value represents an optimal choice. The theoretical analysis and anatomical evidence demonstrate good consistency, supporting the rationale behind the selection of parameters for the directed graph in this study. Therefore, a directed graph with approximately 540 nodes will be used for subsequent experiments.

\subsection*{The probability of a subset of nodes successfully forming a connected subgraph}
Given the connection probability defined in \ref{eq:beconnected}, what is the probability that a subset of nodes (for example, a subset within the directed graph of 540 nodes mentioned above) successfully forms a connected subgraph? Assuming this subset has $M$ nodes, intuition tells us that if $M$ is too small, it indicates that the nodes are sparsely distributed and too far apart from each other. According to \ref{eq:beconnected}, the probability of them forming a connected subgraph would be low. Conversely, if $M$ reaches a sufficient quantity and is uniformly distributed, the probability of them ultimately achieving connectivity would be significantly higher.

Assuming $M$ nodes are uniformly distributed among the aforementioned 540 nodes, they can form a layout similar to a super-pixel partition, where the area of each super-pixel is s=1000$\mu m$*1000$\mu m /M$ . The two nearest neighbor nodes can be approximated to be at the endpoints of the longer diagonal of the super-pixel, so the mathematical expectation of the distance scale between them can be approximated as $\sqrt{s}$. With this neighboring distance value, the connection probability between the two nearest neighbor nodes is $P(be\_connected)_{d=\sqrt{s}}$.

The probability of $M$ nodes being connected equals "1 - the probability of having two isolated branches - the probability of having three isolated branches - ... - the probability of all $M$ nodes being isolated."
For example, to form two isolated branches, let branch 1 contain $M_1$ nodes and branch 2 contain $M_2$ nodes, where $M=M_1+M_2$. The probability that any node in branch 1 does not connect to any node in branch 2 is: $\prod_{x=1}^{M_1} (1-P(be\_connected)_{d_x=J\sqrt{s} }) ^{M_2},J\in\{1,2,3,...,M\} $. This is a product of exponential functions, and when $M\geq20$, the probability becomes very low. Furthermore, if the connection failure results in the formation of three isolated branches (where $M=M_1+M_2+M_3$), the probability is:

This value is also very small (see supplementary information for details). Similarly, the probability of forming more isolated branches can be derived. Therefore, the probability of successfully connecting a set of nodes with a certain size is guaranteed.
\begin{equation}
    \begin{split}
        \prod_{x=1}^{M_1} (1-P(be\_connected)_{d_x=J_1\sqrt{s} }) ^{M_2}\\
        \times \prod_{x=1}^{M_1} (1-P(be\_connected)_{d_x=J_2\sqrt{s} }) ^{M_3}\\
        \times \prod_{x=1}^{M_2} (1-P(be\_connected)_{d_x=J_3\sqrt{s} }) ^{M_3},\\
        J_1,J_2,J_3\in\{1,2,3,...,M\}
    \end{split}
\end{equation}

\subsection*{Experiment on the Relationship Between the Average Unidirectional Connection Saturation of Directed Graph Subgraphs and Subgraph Size}
In the experiment, the directed graph parameters are set to $k$=0.06 and $\lambda$=41. To ensure the statistical significance of the experimental results, the experimental scheme is as follows:

\textbf{1. Generation of Directed Graphs:} Generate 500 nodes uniformly distributed within a $1000 \times 1000$ area. Randomly create 100 directed graphs, all with the same parameters $k$ and $\lambda$, but differing in the details of their topological structure.

\textbf{2. Subgraph Sampling}: For each directed graph, vary the number of nodes from 10 to 300 in increments of 10, generating 40 different scales of subsets. For each scale, randomly select 20 subset samples and compute the average unidirectional connection saturation for each group.

\textbf{3. Observation and Analysis}: After conducting experiments on the 100 directed graphs, observe the average unidirectional connection saturation of subgraphs at different scales of node subsets.
The statistical results of 80,000 directed graphs of different scales are shown in Figure 3(a), with the x-axis representing the scale of the node set and the y-axis representing the average unidirectional connection saturation. From the figure, it can be observed that when the number of nodes exceeds approximately 50, the average unidirectional connection saturation of the node sets of different scales stabilizes around 0.033.

\subsection*{Counting Hamiltonian Cycles in Directed Graphs}

\textbf{Scheme 1:} When the number of nodes $N\leq20$, use a computer to traverse the directed graph and count the number of Hamiltonian circuits:

1. Randomly generate 1000 directed graphs with N uniformly distributed nodes, ensuring that the unidirectional connection saturation of the directed graph is greater than ${\log{N}}/{N}$, with parameters $k$ and $\lambda$ being random.

2. Use the traversal method to calculate how many Hamiltonian circuits exist in one directed graph, then compute the average and compare it with $N!\left({Saturate}_{unidirectional\ connection}\right)^N$.

\textbf{Scheme 2:} When the number of nodes $N\ge20$, use the Monte Carlo method for estimation:
1. Randomly generate 1000 directed graphs with $N$ uniformly distributed nodes, ensuring that the unidirectional connection saturation of the directed graph is greater than $\log{N}/{N}$, with parameters $k$ and $\lambda$ being random.
2. For each selected directed graph, if the number of nodes is $N$, randomly take $M$ different permutations of these $N$ nodes as samples. Then, sequentially check whether these permutations correspond to a traversal order of nodes that forms a Hamiltonian circuit in the directed graph. If $a$ permutations match this condition, then the number of Hamiltonian circuits in that directed graph can be estimated as ${a}/{M}\times(N!)$. 
3. Compare this result with $N!\left({Saturate}_{unidirectional\ connection}\right)^N$.
The experimental results obtained from the two schemes are shown in Figure 3(d), where the measured values approximate the trend calculated using Asaf-Ferber's formula.

\end{document}